\documentclass[12pt]{article}
\usepackage{epsfig}
\newcommand{\noi}{\noindent}
\newcommand{\eq}{\begin{equation}}
\newcommand{\en}{\end{equation}}
\newcommand{\eqa}{\begin{eqnarray}}
\newcommand{\ena}{\end{eqnarray}}

\newcommand{\oOmega}{{\overline \Omega}}

\newcommand{\re}{\mbox{Re}}

\newcommand{\car}{{\cal R}}

\newcommand{\wtmu}{{\widetilde \mu}}
\newcommand{\wtS}{{\widetilde S}}
\newcommand{\tr}{\mbox{Tr}}

\newcommand{\hati}{{\hat i}}
\newcommand{\hatk}{{\hat k}}

\newcommand{\hatmu}{{\hat \mu}}

\newcommand{\hatU}{{\hat U}}
\newcommand{\hatV}{{\hat V}}

\newcommand{\ve}{{\vec e}}

\newcommand{\vb}{{\vec b}}

\newcommand{\vx}{{\vec x}}
\newcommand{\vy}{{\vec y}}

\newcommand{\vE}{{\vec E}}

\newcommand{\vT}{{\vec T}}

\def\be{\begin{equation}}
\def\ee{\end{equation}}
\def\bc{\begin{center}}
\def\ec{\end{center}}
\def\bea{\begin{eqnarray}}
\def\eea{\end{eqnarray}}

\begin{document}

\renewcommand{\theequation}{\arabic{section}.\arabic{equation}}
\renewcommand{\thesection}{\arabic{section}}

\hbox{}
\noindent December 22, 2006 ~~        \hfill JINR E2--2006--166

\vspace{7mm}

\begin{center}
{\Large Disorder parameter and canonical quantization approach in lattice
gauge theory}

\vspace{5mm}

{\bf V.K. Mitrjushkin$^{\rm a,b}$}

\vspace{5mm}

{$^{\rm a}$ {\small\it Joint Inst. for Nuclear Research, 141980 Dubna, Russia
}\\
$^{\rm b}$ {\small\it Inst. of Theor. and Exper. Physics, 117259 Moscow, Russia}\\
}

\end{center}

\vspace{5mm}



\begin{abstract}

This work is dedicated to the derivation of the 'functional' integral
representation of the disorder parameter $\langle\hatmu\rangle$, $\hatmu$
being the the disorder {\it operator}. This derivation resolves the
problem of the choice of the so called modifed action as well as the
problem of the choice of the boundary conditions along the 'time'
direction. The question of the gauge fixing in the functional integral
representation is also discussed.

\end{abstract}

\section{Introduction}\setcounter{equation}{0}

The concept of the disorder parameter \cite{kace} looks very promising
in lattice gauge theories. In particular, a properly defined disorder
parameter $~\langle\hat\mu\rangle~$ gives a possibility to check if the
dual superconductivity of the ground state and monopole condensation
are responsible for the color confinement \cite{thoo}--\cite{niol}.
Last years a number of papers has been dedicated to the (numerical)
study of this hypothesis (see, e.g, papers \cite{adg1} -- \cite{bfkm}
and also papers~\cite{itep1,itep2} where the expectation
value of the monopole creation operator of the Fr\"ohlich-Marchetti
type~\cite{frohl} was shown to be a disorder parameter as well).

\vspace{2mm}

Very often (see, e.g., \cite{adg2}), the disorder parameter $~\langle
\hatmu \rangle~$ is defined as an average of a disorder {\it operator}
$~\hatmu~$ :

$$
\langle \hatmu\rangle \sim \tr \left( \ldots\times\hatmu \right)~;
$$

\noi (for exact definitions see below). On the other hand, in the
functional integral representation (Wilson's approach) the disorder
parameter is written as

\eq
\langle \hatmu\rangle = Z^{-1}\int\! [dU]~
\wtmu(U) \cdot \exp\Big\{ -S_W(U)\Big\}
\equiv Z^{-1}\int\! [dU]~ \exp\{-\wtS(U)\Big\}~,
\label{disorder_wil}
\en

\noi where $~S_W(U)~$ is the standard Wilson action and the partition
function is given by $~Z=\int\! [dU]~\cdot \exp\{-S_W(U)\Big\}~$.

 Therefore, the first question of interest is the connection
between operator $~\hatmu~$ and corresponding functional $~\wtmu(U)~$
(or $\wtS(U)~$).

Another two questions of interest are the following.

\begin{itemize}

\item[-] What kind of boundary conditions along the forth (i.e. imaginary
time) direction should/could one use in the functional integral
representation (note that operator $\hatmu$ is defined in the $3d$
space while functional integral is defined in the $4d$ space) ?

\item[-] Should one fix any gauge in the functional integral ?

\end{itemize}

The next section contains the brief discussion of the canonical
quantization approach on the lattice and the transfer matrix formalism.
The third section is dedicated to the disorder parameter. The last
section is reserved for conclusions.

In what follows the periodic boundary conditions in the spacelike
directions are presumed. The gauge groups is $SU(N)$ , symbol $a_4$
means spacing in the forth direction.


\section{Canonical quantization and transfer matrix}
\setcounter{equation}{0}

Hamiltonian acts in the space $~{\cal H} = \Bigl\{
|\Psi\rangle\Bigr\}~$ where $~\langle\Psi|\Psi\rangle < \infty~$. The
wave function of the state $|\Psi\rangle$ depends on link variables
$~U_{l_s}~$ where $l_s~$ are spacelike links ($l_s \equiv (\vx;\vy)
\equiv (\vx;\vx+{\hat k})~$. Therefore, $~|\Psi\rangle~$ can be
represented in the form

\eq
|\Psi\rangle = \int\!\prod_{l_s}dU_{l_s}~\Psi(U)|U\rangle~,
\en

\noi where $~|U\rangle~$ is the eigenstate of the operators
$~\hatU_{l_s}=\hatU_{\vx;\vy}~$ :
$~\hatU_{l_s}|U\rangle = U_{l_s}|U\rangle~~$, and

\eq
\langle U^{\prime}|U\rangle = \prod_{l_s}
\delta( U^{\prime}_{l_s};U_{l_s})~;
\qquad
\int[dU]~|U\rangle\langle U| ~=~ 1.
\en

\noi Let $~\Omega = \exp\{ i\sum_a\omega^aT^a\} \in SU(N)~$ be some
matrix and for every spacelike link $~l~$ let us define unitary
operators $~R_l(\Omega)~$ :
$~R_l(\Omega) |U\rangle = |U^{\prime}\rangle~;
~~~ R_l({\hat 1}) = {\hat 1}~$,
where $~U^{\prime}_l = \Omega U_l~$ and $~U^{\prime}_{l_1} = U_l~$
for $l_1\ne l~$. It is easy to show,

\eq
R_l(\Omega)\hatU_l R^{\dagger}_l(\Omega) = \Omega^{\dagger} \hatU_l~.
\en

\noi Operators $~R_l(\Omega)~$ can be represented in the form
$~R_l(\Omega) = \exp\Bigl\{ i\sum_a \omega^a_l E^a_l \Bigr\}~$,
where hermitian operators $E^a_l$ have the following commutation
relations

\eq
\Bigl[ E^a,E^b\Bigr] = i f^{abc} E^c~;
\qquad
\Bigl[ E^a,\hatU \Bigr] = -T^a \hatU~;
\qquad
\Bigl[ E^a,\hatU^{\dagger} \Bigr] = \hatU^{\dagger}T^a~.
\en

%

\noi One can establish a connection between $~E^a_l = E^a_{\vx;\vy}~$ and
$E^a_{-l} = E^a_{\vx+{\hat j};-j} = E^a_{\vy;\vx}~$ :

\eq
E^a_{-l} = -\sum_b G^{ab}(U_l) \cdot E^b_l~;
\qquad
G^{ab}(U_l) = 2\tr (U_l T^a U_l^{\dagger}T^b)~,
\en


\noi Evidently,

\eq
\Bigl[ E^a_{-l},E^b_{-l}\Bigr]
= -if^{abc} E^c_{-l} ~;
\qquad
\Bigl[ E^a_{-l},\hatU_l \Bigr] =
\hatU_l T^a ~;
\qquad
\Bigl[ E^a_{-l},\hatU^{\dagger}_l \Bigr]
= -T^a\hatU^{\dagger}_l~.
\en

\noi The generator $\Gamma^a(\vx)$ of the gauge transformation in $\vx$
and the operator $\car(\Omega)$ of the gauge transformation are given
by

\eq
\Gamma^a(\vx) \equiv \sum_{k=1}^3 \Bigl(E^a_k(\vx)+E^a_{-k}(\vx)\Bigr)~;
\qquad
\car(\Omega) = \prod_{\vx} e^{i\sum_a\omega^a(\vx)\Gamma^a(\vx)}~,
         \label{gauge_gener}
\en

\noi and

\eq
\car(\Omega)\hatU_k(\vx)\car^{\dagger}(\Omega)
= \Omega^{\dagger}(\vx)\,\hatU_k(\vx)\,\Omega(\vx+\ve_k)~.
\en

\vspace{2mm}

\noi The vacuum state $~|0\rangle~$ is supposed to be gauge invariant

\eq
\car(\Omega)|0\rangle = |0\rangle~;
\qquad
\Gamma^a(\vx)|0\rangle = 0~.
\en

\noi The second equation means that the electric flux equals to zero if
there are no sources (Gauss law).
It is easy to show that operator

\eq
P_0 = \int\! [d\Omega]~\car(\Omega) \equiv \prod_{\vx}
\int\! d\Omega(\vx)~\exp\left\{ i\sum_a
\omega(\vx)\Gamma^a(\vx) \right\}~.
\en

\noi is the projection operator on the gauge invariant states.

\vspace{3mm}

Let $H$ be a Hamiltonian (to be defined later). Then the partition
function $~Z~$ is

\eq
Z = \tr \left( \exp\Bigl\{ -\frac{1}{T}H \Bigr\}\right)_{colorless}~;
\qquad T\equiv \frac{1}{a_4N_4}~,
\en

\noi where $T$ is the temperature and the trace is defined on some
colorless space of states. By definition, the transfer matrix $\hatV$
is given by $~\hatV = \exp\{-a_4H \}~$.
Therefore,

\eq
Z = \tr \left( \hatV^{N_4} \right)_{colorless}
~\equiv~ \tr \left( \hatV^{N_4} P_0\right)~.
\en

\noi Let $~|\Psi_k\rangle~$ be eigenstates of the transfer matrix
$\hatV~$ with eigenvalues $~\lambda_k~$

\eq
\hatV|\Psi_k\rangle = \lambda_k|\Psi_k\rangle~;
\qquad
\lambda_k \equiv e^{-a_4E_k}~,
\en

\noi and $~E_k~$ are eigenvalues of the Hamiltonian. One can choose

\eq
E_0 < E_1\le E_2\le \ldots~.
\en

\noi Vacuum $~|\mbox{vac}\rangle~$ is the eigenstate with the lowest
$~E_0~$ : $~|\mbox{vac}\rangle \equiv |\Psi_0\rangle~$, and it is
supposed to be nondegenerate. The partition function $Z$ can be
represented as

\eq
Z= \tr \left( \hatV^{N_4} P_0\right)
~\equiv~ \sum_{k\ge 0} \langle \Psi_k |\hatV^{N_4} P_0
|\Psi_k\rangle
~=~ \sum_{k\ge 0} \exp\left\{-\frac{1}{T}E_k\right\} \cdot
\langle\Psi_k |P_0|\Psi_k\rangle~.
\en

\noi Every eigenstate can be expanded as follows

\eq
|\Psi_k\rangle = \int\! [dU]~\Psi_k(U)|U\rangle~.
\en

\noi The set of the eigenstates $~\{ |\Psi_k\rangle \}~$is supposed to
be complete

\eq
\sum_k \Psi_k^{\ast}(U^{\prime})\Psi_k(U) = \delta(U;U^{\prime})~.
\en

\noi Then

\eqa
Z &=& \sum_{k\ge 0} \langle \Psi_k |\hatV^{N_4} P_0|\Psi_k\rangle
~=~ \sum_{k\ge 0} \int\! [dU][dU^{\prime}]
~\Psi_k^{\ast}(U^{\prime})\Psi_k(U)\cdot \langle U^{\prime}|\hatV^{N_4}
P_0|U\rangle
\nonumber \\
\nonumber \\
&=& \int\! [dU]~\langle U|\hatV^{N_4}P_0|U\rangle~.
\ena

\noi Evidently, in the zero temperature limit the main contribution to
the partition function $~Z~$ comes from the vacuum eigenstate

\eq
Z = \langle\mbox{vac}|\exp\Bigl\{ -\frac{1}{T}H \Bigr\}|\mbox{vac}\rangle~;
\qquad T\equiv \frac{1}{a_4N_4}\to 0~.
\en

\vspace{3mm}

The partition function $~Z~$ defined above presumes the time
periodicity

\eq
U_{\mu}(\vx;x_4=L_4)=U_{\mu}(\vx;x_4=0)~.
\en

\noi By applying the gauge transformation $~\Omega(\vx;x_4)~$

\eq
\Omega(\vx;x_4) = \left\{
\begin{array}{rl}
U_4^{\dagger}(\vx;L_4-a_4)~;
& ~~x_4 = L_4-a_4~ ;  \\
\Bigl( U_4(\vx;L_4-2a_4)U_4(\vx;L_4-a_4)\Bigr)^{\dagger}~;
& ~~x_4 = L_4-2a_4~ ; \\
\ldots
& \ldots              \\
\Bigl( U_4(\vx;a_4) U_4(\vx;2a_4)\ldots U_4(\vx;L_4-a_4)\Bigr)^{\dagger}~;
& ~~x_4 = a_4~,
\end{array}\right.
\en

\noi one obtains $~U_4(\vx;x_4)=1~$ at $~x_4\ne 0~$~. Let us define
$~\Lambda(\vx) \equiv U_4(\vx;0)~$, and

\eq
U_k^{\Lambda}(\vx;0) = \Lambda^{\dagger}(\vx)U_k(\vx;0)
\Lambda(\vx+{\hat e}_k)~.
\en

\noi Evidently, $~\tr\, U_{P_s}^{\Lambda}(x_4) = \tr\, U_{P_s}(x_4)~~$.
Defining functions $~W_E,W_M~$ :

\eqa
W_E(0) &=& \frac{2N a}{g^2_ta_4}\sum_{k;\vx}\left[ 1 - \frac{1}{N}
\re\tr \Bigl(U_k^{\Lambda}(\vx;0)U_k^{\dagger}(\vx;a_4)\Bigr)\right]~;
\hspace{19mm} x_4= 0~;
\nonumber \\
\nonumber \\
W_E(x_4) &=& \frac{2N a}{g^2_ta_4}\sum_{k;\vx}\left[ 1 - \frac{1}{N}
\re\tr \Bigl(U_k(\vx;x_4)U_k^{\dagger}(\vx;x_4+a_4)\Bigr)\right]~;
\qquad x_4\ge a_4~;
\nonumber \\
\nonumber \\
W_M(x_4) &=& \frac{2N a_4}{g^2_s a}
\sum_{P_s} \left( 1 -\frac{1}{N} \re\tr U_{P_s}(x_4) \right)~,
\ena

\noi one obtains

\eq
S(U) = \sum_{x_4=0}^{L_4-a_4} \left\{ W_E(x_4) + \frac{1}{2}\Bigl[
W_M(x_4)+W_M(x_4+a_4)\Bigr]\right\}~.
\en

\noi Let us introduce the (pure gauge) transfer matrix $~\hatV_G~$ with
matrix elements

\eq
\langle U(x_4+a_4)|\hatV_G |U(x_4)\rangle = C_0^{-1}\cdot\exp\left\{
- W_E(x_4)-\frac{1}{2}\Bigl[W_M(x_4)+W_M(x_4+a_4)\Bigr]\right\}~,
                \label{matrix_elem}
\en

\noi where $~C_0~$ is some constant. Then

\eq
Z = \int\![d\Lambda]\int\!\prod_{x_4=0}^{L_4-a_4}[dU(x_4)]
~\exp\left\{-\sum_{x_4} \Bigl[ W_E(x_4)+ W_M(x_4) \Bigr]\right\}
= C_0^{N_4}\cdot\tr \left( \hatV^{N_4}_G P_0\right)~,
\en

\noi where

\eq
[d\Lambda] = \prod_{\vx}d\Lambda(\vx)~;
\qquad
[dU(x_4)] = \prod_{\vx}\prod_k dU_k(\vx;x_4)~,
\en

\noi and $P_0$ is the projection operator on the colorless state
defined above.

Let us note that in the finite volume one obtains the standard
Wilson action only in the case of the periodic boundary condition, i.e.
when $~W_M(L_4)=W_M(0)~$.


\section{Disorder parameter $\langle \hatmu\rangle$}
\setcounter{equation}{0}

For every spacelike link  $~l_s=(\vx;k)~$ let us define matrices

\eq
\oOmega_{l_s} \equiv \exp\Bigl\{ i\vb_{l_s} \vT\Bigr\} \in SU(N)~,
\en

\noi where $~\vb_{l_s}~$ are any parameters (not necessarily connected
with monopoles). Following the Pisa approach one can define the disorder
operator

\eq
\hatmu(\vb) = \exp\Bigl\{ i\sum_{l_s} \vb_{l_s} \vE_{l_s}\Bigr\}~.
\en

\noi Then the disorder parameter $\langle\hatmu\rangle$ is given by

\eq
\langle\hatmu\rangle = \frac{1}{Z}Z(\oOmega)~;
\qquad
Z(\oOmega) = \tr \left( \prod_{l_s}R_{l_s}(\oOmega_{l_s})\,
\hatV^{N_4}_G P_0\right)~,
\en

\noi where operators $~R_{l_s}(\oOmega_{l_s})~$, $\hatV_G$ and $P_0$
have been defined in the previous section. Repeating the procedure of
the previous section, one obtains in the temporal gauge (tg) :

\eq
Z(\oOmega) = \int\![dU]_{tg} ~e^{ -S(U;\oOmega)}~;
\qquad
[dU]_{tg} \equiv \prod_{\vx}dU_4(\vx,0)\prod_{x_4=0}^{L_4-a}
\prod_{\vx k} dU_k(\vx;x_4)~,
          \label{part_tg}
\en

\noi where

\eqa
S(U;\oOmega) &=& S_W(U)+\delta S_E(U;\oOmega) + \delta S_M(U;\oOmega)~;
\nonumber \\
\nonumber \\
\delta S_E
&=& \frac{2}{g^2} \sum_{k;\vx}\re\tr \Bigl[ (1-\oOmega_{\vx;k})\cdot
U_k(\vx;L_4-a) U_k^{\dagger}(\vx;0) \Bigr]~;
           \label{delta_S_E}
\nonumber \\
\\
\delta S_M &=&
\frac{1}{g^2} \sum_{i<k}\sum_{\vx} \re\tr \left[ U_{\vx;i}U_{\vx+\hati;k}
U_{\vx+\hatk;i}^{\dagger}U_{\vx;k}^{\dagger}
- U_{\vx;i}^{\prime}\cdot
U_{\vx+\hati;k}^{\prime} \cdot U_{\vx+\hatk;i}^{\prime,\dagger} \cdot
U_{\vx;k}^{\prime,\dagger} \right](x_4=0) ~.
           \label{delta_S_M}
\nonumber
\ena

\noi and $~U^{\prime}_{l_s} = \oOmega^{\dagger}_{l_s}U_{l_s}~$.
To get rid of the temporal gauge one can perform successively the
change of variables $~U \to U^{\prime\prime}~$ :

\eqa
U_k(\vx;x_4) &=& U_4^{\dagger}(\vx;x_4) U_k^{\prime\prime}(\vx;x_4)
U_4(\vx+\hatk;x_4)~;
\\
U_4(\vx;x_4-a) &=& U_4^{\prime\prime}(\vx;x_4-a)U_4(\vx;x_4)~
\ena

\noi at $x_4=a$, then at $x_4=2a$, $~\ldots~$, then at $x_4=L_4-a~$.
Finally, one arrives at

\eqa
\langle\hatmu(\oOmega)\rangle &=& \frac{1}{Z}\int\![dU]~
e^{-\wtS(U;\oOmega)}~;
\qquad
\wtS(U;\oOmega) = S_W(U)+\delta S_E(U;\oOmega)~;
       \label{hatmu_1}
\\
\nonumber \\
\delta S_E(U;\oOmega) &=& \frac{2}{g^2} \sum_{k;\vx}\re\tr \Bigl[
(1-\oOmega_{\vx;k}) \cdot U_{\vx;k}(0)U_{\vx+\hatk;4}(0)
U_{\vx;k}^{\dagger}(a)U_{\vx;4}^{\dagger}(0)
\Bigr]~.
       \label{hatmu_2}
\ena

Equations (\ref{hatmu_1}) and (\ref{hatmu_2}) define the desired --
functional -- representation of the order parameter
$~\langle\hatmu\rangle~$.


\section{Summary}\setcounter{equation}{0}

Finally, let us summarize.
This paper is dedicated to the discussion of some formal aspects of the
functional integral representation of the disorder parameter in lattice
gauge theories. No special choice of the fields $\vb_{l_s}$ has been used.

\vspace{2mm}

The main results are the following.

\begin{itemize}

\item[{\bf 1)}] The functional integral representation of the disorder
parameter $~\langle\hatmu\rangle~$ has been derived in the canonical
quantization approach. This derivation resolves the problem --
discussed in paper \cite{adg1} -- of the (unambiguous) choice of the
modifed action $\wtS(U;\oOmega)$ on the lattice.

The representation obtained for the disorder parameter $~\langle \hatmu
\rangle~$ is given in eq.'s~(\ref{hatmu_1}) and (\ref{hatmu_2}) and
coincides with that used, say, in paper \cite{adg6}.

\item[{\bf 2)}] As a byproduct, this derivation establishes the
boundary conditions along the imaginary time direction : they have to
be periodic.

The choice of, say, $C^{\ast}$--periodic boundary conditions (see,
e.g., \cite{adg6}) is not compatible with the canonical quantization
approach.

\item[{\bf 3)}] Also, this derivation resolves the problem of the choice
of the gauge in the functional integral representation for $~\langle
\hatmu \rangle~$ : no gauge fixing  is needed.

This is in agreement with the numerical results of the paper
\cite{adg6} where the so called RAP (i.e. no gauge fixing) has been
also used.

\end{itemize}

\section*{Acknowledgments}

This work has been supported by the RFBR grant 05-02-16306,
DFG--RFBR grant 06-02-04014  and Heisenberg--Landau program.


\begin{thebibliography}{99}
\newcommand{\prd}[1]{Phys.~Rev.~{\bf D#1}\ }
\newcommand{\prb}[1]{Phys.~Rev.~{\bf B#1}\ }
\newcommand{\plb}[1]{Phys.~Lett.~{\bf B#1}\ }
\newcommand{\npb}[1]{Nucl.~Phys.~{\bf B#1}\ }
\newcommand{\prl}[1]{Phys.~Rev.~Lett.~{\bf #1}\ }
\newcommand{\pr}[1]{Phys.~Rep.~{\bf #1}\ }
\newcommand{\ap}[1]{Ann.~Phys.~{\bf #1}\ }
\newcommand{\cmp}[1]{Commun.~Math.~Phys.~{\bf #1}}
\newcommand{\rmp}[1]{Rev.~Mod.~Phys.~{\bf #1}}
\newcommand{\ptp}[1]{Prog.~Theor.~Phys.~{\bf #1}}




\bibitem{kace} L.P.~Kadanoff and H.~Ceva, \prb{3} (1971) 3918.
\bibitem{thoo} G.~'t~Hooft, in {\em High Energy Physics}, Proceedings of
the EPS International Conference, Palermo 1975, ed. A.~Zichichi,
Editrice Compositori, Bologna 19976.
\bibitem{mand} S.~Mandelstam, \pr{23} (1976) 245.
\bibitem{niol} H.B.~Nielsen and P.~Olesen, \npb{61} (1973) 61.
\bibitem{adg1} L.~Del Debbio, A.~Di Giacomo and G.~Paffuti,
\plb{349} (1995) 513.
\bibitem{adg2} A.~Di Giacomo and G.~Paffuti, \prd{56} (1997) 5816.
\bibitem{adg3} L.~Del Debbio, A.~Di Giacomo, G.~Paffuti and
P.~Pieri, \plb{355} (1995) 255.
\bibitem{adg4} A.~Di Giacomo, B.~Lucini, L.~Montesi and G.~Paffuti,
\prd{61} (2000) 034503.
\bibitem{adg5} A.~Di Giacomo, B.~Lucini, L.~Montesi and G.~Paffuti,
\prd{61} (2000) 034504.
\bibitem{adg6} J.M.~Carmona, M.~D'Elia, A.~Di Giacomo, B.~Lucini
and G.~Paffuti, \prd{64} (2001) 114507.
\bibitem{adg7} A.~Di Giacomo, hep-lat/0206018 (2002).
\bibitem{adg8} M.~D'Elia, A.~Di Giacomo, B.~Lucini, G.~Paffuti and
C.~Pica, \prd{71} (2005) 114502.

\bibitem{bbem} A. Barresi, G. Burgio, M. D'Elia and M. Mueller-Preussker,
\plb{599} (2004) 278.
\bibitem{bfkm} G. Burgio, M. Fuhrmann, W. Kerler and M. Mueller-Preussker,
hep-lat/0610097 (2006).

\bibitem{itep1}   M.~N.~Chernodub, M.~I.~Polikarpov and A.~I.~Veselov,
\plb{399} (1997) 267;
Nucl. Phys. Proc. Suppl. {\bf 49} (1996) 307;
\bibitem{itep2}   V.~A.~Belavin, M.~N.~Chernodub and M.~I.~Polikarpov,
  JETP Lett. {\bf 83} (2006) 308.
\bibitem{frohl} J.~Frohlich and P.~A.~Marchetti,
\cmp{112} (1987) 343.

\end{thebibliography}
\end{document}